\documentclass[runningheads]{llncs}
\usepackage{graphicx}
\usepackage{amssymb}
\usepackage{amsmath}
\usepackage{multirow}
\usepackage{authblk}
\begin{document}
\title{LoCI-DiffCom: Longitudinal Consistency-Informed Diffusion Model for 3D Infant Brain Image Completion}
%
%
\author{Zihao Zhu$^{1,\dagger}$ \and Tianli Tao$^{1,\dagger}$ \and Yitian Tao$^{1,\dagger}$ \and Haowen Deng$^{1}$ \and Xinyi Cai$^{1}$ \and Gaofeng Wu$^{1}$ \and Kaidong Wang$^{1}$ \and Haifeng Tang$^{1}$ \and Lixuan Zhu$^{1}$ \and Zhuoyang Gu$^{1}$ \and Jiawei Huang$^{1}$ \and Dinggang Shen$^{1,2,3}$ \and Han Zhang$^{1,*}$\thanks{$\dagger$ indicates co-first authors and $*$ indicates corresponding author (e-mail: zhanghan2@shanghaitech.edu.cn).}}


\institute{School of Biomedical Engineering, ShanghaiTech University, Shanghai, China \and Shanghai United Imaging Intelligence Co., Ltd, Shanghai, China \and Shanghai Clinical Research and Trail Center, Shanghai, China}
%

\makeatletter
\def\thanks#1{\protected@xdef\@thanks{\@thanks
        \protect\footnotetext{#1}}}
\makeatother

\maketitle              
\begin{abstract}
The infant brain undergoes rapid development in the first few years after birth. Compared to cross-sectional studies, longitudinal studies can depict the trajectories of infants’ brain development with higher accuracy, statistical power and flexibility. However, the collection of infant longitudinal magnetic resonance (MR) data suffers a notorious dropout problem, resulting in incomplete datasets with missing time points. This limitation significantly impedes subsequent neuroscience and clinical modeling. Yet, existing deep generative models are facing difficulties in missing brain image completion, due to sparse data and the nonlinear, dramatic contrast/geometric variations in the developing brain. We propose LoCI-DiffCom, a novel Longitudinal Consistency-Informed Diffusion model for infant brain image Completion, which integrates the images from preceding and subsequent time points to guide a diffusion model for generating high-fidelity missing data. Our designed LoCI module can work on highly sparse sequences, relying solely on data from two temporal points. Despite wide separation and diversity between age time points, our approach can extract individualized developmental features while ensuring context-aware consistency. Our experiments on a large infant brain MR dataset demonstrate its effectiveness with consistent performance on missing infant brain MR completion even in big gap scenarios, aiding in better delineation of early developmental trajectories.


\end{abstract}
\section{Introduction}
The brains of human infants undergo dramatic morphometric and geometric changes during early infancy. The total cerebral volume increases from about 30\% to 80\% of the adult size during the first two years after birth\cite{gilmore2018brainDevelopment,knickmeyer2008structural}.
Besides global changes, local brain areas evolve more significantly\cite{knickmeyer2008structural,holland2014structural}, laying foundations for emerging cognitive and learning abilities\cite{paterson2006development}. Recent advancements more and more rely on the use of longitudinal magnetic resonance imaging (MRI) to characterize growth trajectories\cite{howell2019unc,soh2014cohort}. Compared to cross-sectional data, longitudinal MRI can unravel developmental trajectories, especially individual differences in these curves, with elevated accuracy, statistical power, and analytic flexibility\cite{kraemer2000can}. However, longitudinal infant MRI faces enormous challenges due to poor cooperation during scanning, heavy imaging noise and artifacts\cite{knickmeyer2008structural}, even subject drop-out during follow-up stages, resulting in missing data\cite{zhang2018infant}. Commonly, the age interval of two data from the same infant is too large to reveal dynamic and nonlinear developmental changes in-between. Moreover, the contrast of longitudinal infant brain MRI changes dramatically due to immature myelination\cite{hazlett2012brain}, further complicating data completion. The field calls for high-fidelity image completion with accuracy for small brain structures that undergo larger changes.

Deep generative models, such as generative adversarial networks (GANs) and diffusion probabilistic models (DPMs)\cite{ho2020ddpm,song2020denoising}, have achieved significant success in the field of image generation. DPMs have particularly shown superiority in generating 3D medical images with rich details\cite{dorjsembe20223DBrain,pinaya2022brain_ldm,peng2022cdpm_Brain}. The pioneering studies have attempted to use DPMs for longitudinal image completion\cite{kim2022diffusiondeformable,guo2024cas,yoon2023sadm}, but this task remains a cutting-edge challenge for infant brain. For instance, generating deformation fields\cite{kim2022diffusiondeformable} relies on largely uniform deformation assumption, which does not always hold true for infant brain MRI. Another longitudinal MRI completion study relied on single guidance image\cite{guo2024cas}, which might cause large distortions in the infant scenario. Multiple guiding images will offer a better-controlled condition for DPMs by explicitly using multiple past time points to predict future data\cite{yoon2023sadm}. However, the sequence-aware transformer used was borrowed from a video-based vision transformer\cite{arnab2021vivit}, only suitable for extracting simple temporal features from natural video frames but could fail in more complicated infant MRI completion. Learning longitudinal sequence in early brain development is not as simple as that used in video frame completion, an efficient yet powerful algorithm that can integrate spatiotemporal semantic information from very sparse sequences is urgently needed.

This paper presents Longitudinal Consistency-Informed Diffusion model for infant brain image Completion (LoCI-DiffCom). This novel algorithm can provide adaptive guidance to constrain a conditional DPM for generating high-fidelity missing infant brain MRI data with any paired preceding and subsequent time points of any interval. As the missing data bares dramatic temporal variations and large individual variability, we introduce a longitudinal consistency-informed module that fuses two time-point data to achieve context-aware consistency for carefully guiding DPM-based generation. As enormous spatiotemporal information is involved, to adaptively adjust the significance of various semantic features across spatial and channel domains, we incorporate light-weighted channel-spatial attention\cite{liu2021gam} into DPM. Our method on completing the Baby Connectome Project (BCP) dataset\cite{howell2019unc} demonstrates its efficacy in a sparse sequence scenario.

\section{Method}
\begin{figure}[t]
\centering
\includegraphics[width=1\textwidth]{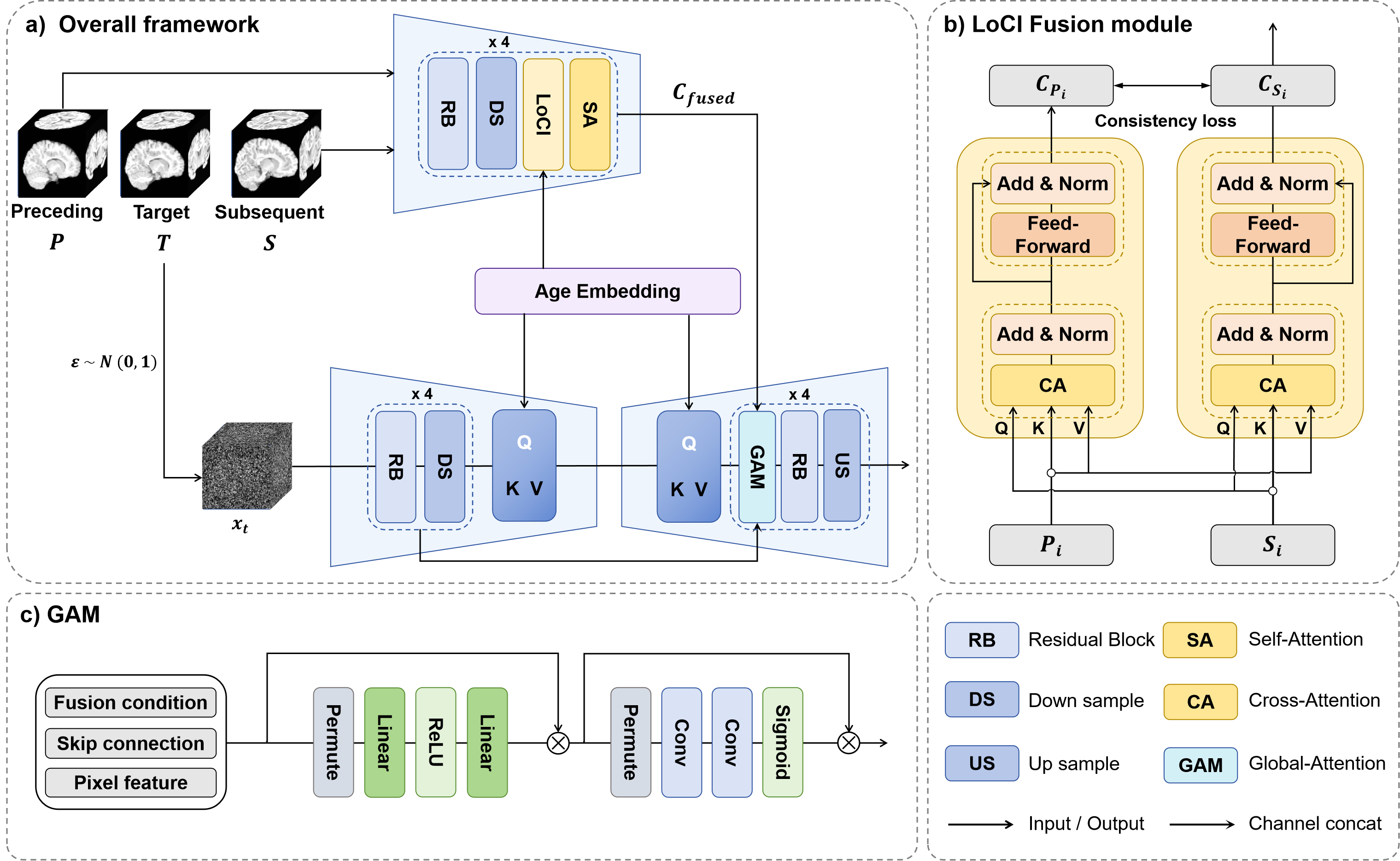}
\caption{The overall architecture of our proposed LoCI-DiffCom (a) with Longitudinal Consistency-Informed (LoCI) Fusion module (b) and a global attention mechanism (GAM) (c).}
\label{Fig:fusion_pipline}
\end{figure}

Fig. \ref{Fig:fusion_pipline}a illustrates the architecture of our LoCI-DiffCom. Specifically, a LoCI module, detailed in Fig. \ref{Fig:fusion_pipline}b, is designed to fuse the preceding and subsequent data to form conditions $C_{fused}$ with context consistency for guiding diffusion model generation. Then, $C_{fused}$ is informed into the DPM via a global attention mechanism (GAM, Fig. \ref{Fig:fusion_pipline}c) for modulating the importance of various semantic information in the denoising network, enabling the better decoding.




\subsection{Longitudinal Consistency-Informed Fusion Module}
In conditional DPMs, the generated image largely depends on the congruence and association between the conditions and the target. For infant early brain development, the condition image may differ significantly from the target image due to infants’ rapid developmental rate. 
Therefore, to accurately guide longitudinal image generation, we design a LoCI fusion module to better integrate images collected preceding ($P$) and subsequent ($S$) to the missing time point, resulting in semantic features $C_{fused}$ with high relevance to the target. This strategy also maximally utilizes existing data for high-fidelity image completion.

In Fig. \ref{Fig:fusion_pipline}a. $P$ and $S$ are initially encoded to extract their semantic representations. The encoder consists of $n$ residual blocks and $n$ two-fold downsampling to get $n$ pairs of encoded features of varied sizes, denoted as $P_i$ and $S_i$. In order to extract the individual's developmental characteristics by integrating the semantic features of preceding and subsequent time points, we input each set of $P_i$ and $S_i$ into the LoCI module (detailed in Fig. \ref{Fig:fusion_pipline}b). LoCI has a transformer-based architecture with multi-head cross-attention, consisting of three Transformer encoders. The initially encoded features sequentially pass through the three cross-attention transformers with their respective $Q, K, V$ derived from fully connected linear transformations. By switching $Q$ to perform cross-attention, followed by layer normalization and feed-forward network, the features from the preceding time point gradually exchange information with those from subsequent time point, with their commonality enhanced. After LoCI fusion, the initial features turn into fused features with context-aware consistency $C_{Pi}$ and $C_{Si}$. To achieve more accurate feature fusion, we minimize the Mean Squared Error (MSE) between $C_{Pi}$ and $C_{Si}$, aiming to reach common characteristics that represents individual's unique developmental traits with high relevance to the target. The loss function, $L_{LoCI}$, can be formulated as:
\begin{equation}
       L_{LoCI} = MSE(C_P, C_S) 
       = \frac{1}{n}\sum_{i=1}^n {(C_{Pi} - C_{Si})^2},
    \label{formula:  eq4}
\end{equation}
where $n$ is the number of LoCI fusion modules. The output of the LoCI module is $C_{Si}$, which has already integrated information from preceding time point and is then fed into a final Transformer encoder with self-attention, resulting in fused conditions $C_{fused}$, as the guidance for the denoising network.

\subsection{Hybrid Attention Mechanism in the DPMs}
In the implementation of conditional DPMs, it is straightforward that the individual guidance image and the age information are encoded and then simply concatenated, or directly added to the latent space, prior to the denoising process. However, this approach may obfuscate features with different semantics. Instead of doing so, we adopt a hybrid attention mechanism in the implementation of LoCI-DiffCom. That is, we embed the fused image feature $C_{fused}$ and the age information $x_{mo}$ using global attention and cross-attention, respectively. As illustrated in Fig. \ref{Fig:fusion_pipline}c, the global attention mechanism (GAM) comprises a lightweight channel attention mechanism and spatial attention. In channel attention, a 4D permutation merges spatial dimensions and swaps their order with the channel dimension, followed by a two-layer multi-layer perceptron (MLP) with a channel reduction ratio $r$ to extract a channel attention map. In spatial attention, the original dimension permutation is first restored, followed by two convolutional layers focusing on extracting an attention map for the spatial dimension. This attention is capable of learning significant global information by traversing dimensions across spatial and channel domains. For the age information of the target image $x_{mo}$, we embed it as textual features and utilize cross-attention between pixel features at both the encoder and decoder of the denoising network.

\subsection{Model Training}
The objective function of the conditional DPM is defined as follows:
\begin{equation}
    \begin{aligned}
        L_{diff} = E[||\epsilon - \epsilon_{\theta}(x_{t};t, x_{mo}, C_{fused})||^{2}],
    \end{aligned}
    \label{formula:  eq6}
\end{equation}
where $\epsilon$ represents a random Gaussian distribution, $x_{t}$ is the noised target image, $t$ represents the number of time steps for adding noise. Given that $C_{fused}$ is the input to the DPM, the objective function of the DPM $L_{diff}$ will optimize the parameters of the LoCI fusion module $\theta_{LoCI}$. However, $L_{fusion}$ does not affect the parameters of diffusion $\theta_{diff}$. Our network employs an end-to-end training approach, allowing simultaneous optimization of both parts, formulated as:
\begin{equation}
    \begin{aligned}
        L = L_{diff} + \lambda L_{fusion}
    \end{aligned}
    \label{formula:  eq7}
\end{equation}
Given the convergence rate of the LoCI module being greater than that of the denoising network, $\lambda$ can be selected to modulate the extent of optimization for the LoCI module.

\section{Experiments and Results}
\subsection{Dataset and Implemention}
\subsubsection{BCP Dataset}In our study, we utilized a longitudinal infant MRI dataset from Baby Connectome Project (BCP)\cite{howell2019unc}, comprising a total of 170 infants aged 0-26 months, with 655 T1-weighted structural MR scans. After rigorous quality control, we randomly selected 584 scans from 154 infants for training and 71 scans from 16 different infants for testing. All data were preprocessed according to a standard procedure\cite{li2013mapping}. Considering the computational cost, input images were resampled to 2$\times$2$\times$2 $mm^3$.

\subsubsection{Implementation Details}
Our model is implemented using PyTorch\cite{paszke2019pytorch} trained on an Nvidia A100 GPU with memory of 80 GB. For diffusion, the noise level is set from $10^{-4}$ to $5 \times 10^{-3}$ linearly with 1000 steps. Adam optimization\cite{kingma2014adam} is used with a learning rate of $2 \times 10^{-4}$. During inference, denoising is performed with 80 skip steps across 1000 steps. The model in our experiment has $n = 4$ LocI modules. $\lambda = 0.6$ is selected to modulate the optimization for LoCI. Channel reduction ratio $r = 4$ in GAM.



\begin{figure}[!b]
\centering
\includegraphics[width=\textwidth]{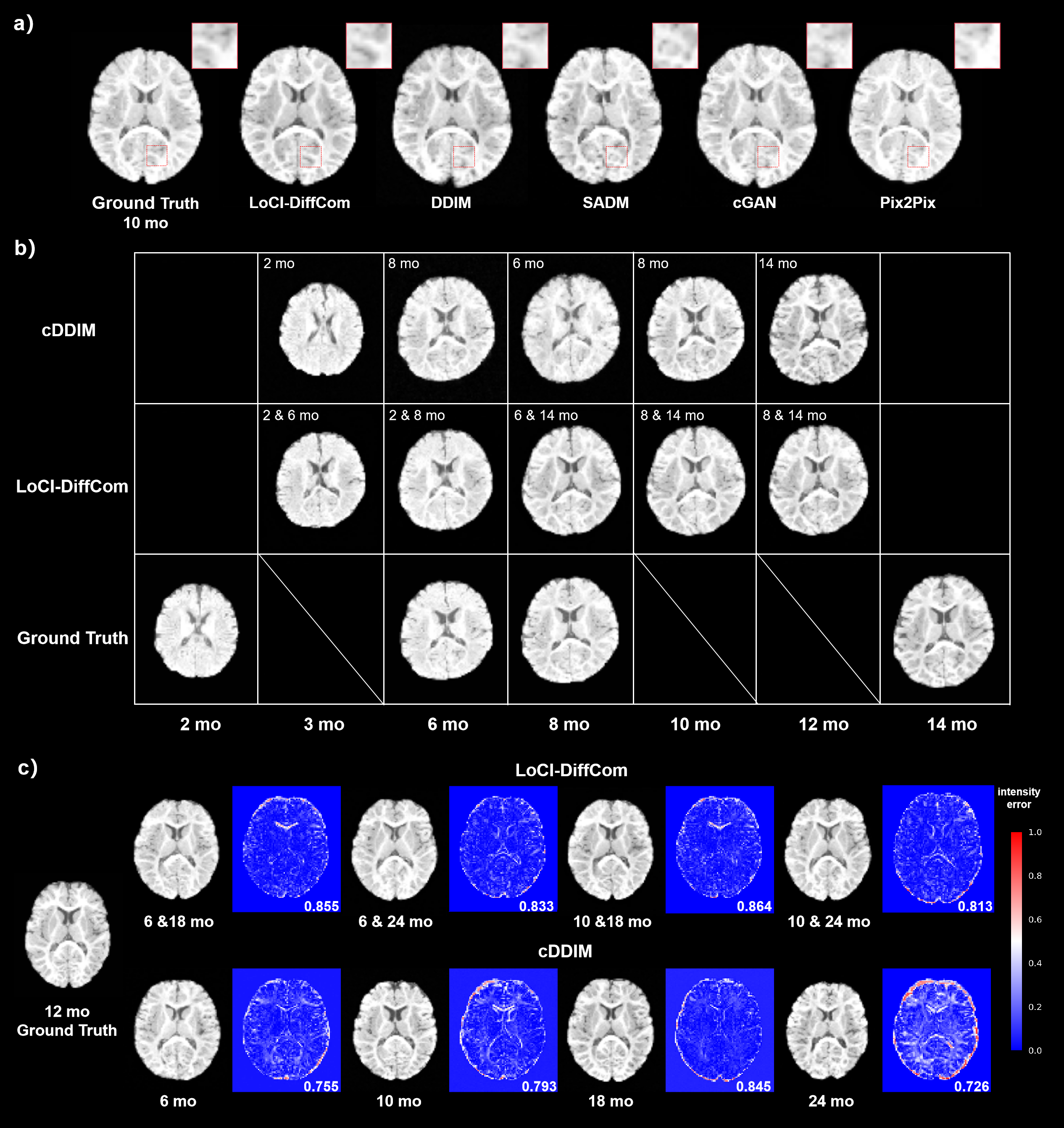}
\caption{Qualitative comparison between baseline methods and our proposed LoCI-DiffCom (a). Comparison of longitudinally generative performance, with the text on the top-left representing conditional images utilized (b). Visual comparison and error maps of different longitudinal consistency guidance. The bottom-right of the error map represents SSIM score (c).}
\label{Fig:result}
\end{figure}

\subsection{Results}
\subsubsection{Evaluation Metrics and Baseline Methods}
We choose several state-of-the-art methods as baseline models for comparison: 1) GAN-based methods: Conditional GAN\cite{mirza2014conditional} and Pix2Pix GAN\cite{isola2017p2p}, which use a UNet-based GAN model to synthesize the missing image given a reference image as the condition, as well as 2) Diffusion-based methods: Conditional DDIM\cite{guo2024cas} and SADM\cite{yoon2023sadm}, which use a diffusion model with longitudinal sequence as conditions. The methods are compared using peak signal-to-noise ratio (PSNR) and structural similarity index measure (SSIM). Furthermore, we evaluate the fidelity of the generated images. Specifically, we employ infant-dedicated brain MRI segmentation\cite{shi2022deep}\cite{zhang2018longitudinally}\cite{shi2010neonatal} to parcellate brain gray matter and white matter using our generated images and use Dice coefficient to evaluate the segmentation performance. A higher Dice score indicates that the generated image is more reasonable.

\subsubsection{Qualitative and Quantitative Comparisons}
The GAN-based method and conditional DDIM used a single MRI at 7 months of age as guidance. SADM was guided by several preceding MRIs at 4 and 7 months of age. Our LoCI-DiffCom utilized the images from both 7 and 14 months as guidance. As shown in Fig. \ref{Fig:result}a, our method outperformed other methods in terms of image details and preserved more individual-specific information. Quantitative evaluations are summarized in Table. \ref{Table:1}. LoCI-DiffCom outperformed GAN-based methods by 5\%-10\% and was slightly better than the diffusion-based methods. It is worth noting that SADM’s results, with preceding time points’ images as guidance, were even worse than those using only one guidance image, largely due to the prominent structural diversity in early development.

Since conditional DDIM performed best among the baseline methods, we visually compare its generated longitudinal sequence with ours. The age(s) of guidance image(s) is shown in the top left corner of each sub-figure in Fig. \ref{Fig:result}b. Our model has better longitudinal consistency, showing a gradual grow-up trend. However, the brain size of the images generated by the conditional DDIM varied in an unreasonable manner.

\begin{table}[!t] 
\centering
\setlength{\tabcolsep}{2mm}
\caption{Quantitative comparison between baselines and our proposed LoCI-DiffCom.}
\begin{tabular}{cccccc}
\hline
\multirow{2}{*}{Method} & \multirow{2}{*}{PSNR} & \multirow{2}{*}{SSIM} & \multicolumn{2}{c}{Dice} \\ \cline{4-5} 
                  &                       &                       & WM     & GM     \\ \hline
cGAN         &23.73                  &0.794                  & 0.642  & 0.629  \\ 
Pix2Pix         &24.01                 &0.796                  & 0.643  & 0.622   \\ 
DDIM           &24.07                 &0.798                  & 0.646  & 0.627   \\
SADM           &23.57                 &0.781                  & 0.569  & 0.587   \\ 
LoCI-DiffCom  &\textbf{25.52}        &\textbf{0.845}         &\textbf{0.656}   &\textbf{0.650} \\
\hline
\end{tabular}
\label{Table:1}
\end{table}

\subsubsection{Ablation Study}
We examined the impact of the number of our proposed LoCI modules on the model’s effectiveness and quantitatively compared the image quality and fidelity. As shown in Table. \ref{Table:2}, LoCI ($\times 4$) achieves a 3\% improvement in SSIM compared to LoCI ($\times 1$). We also investigated the effectiveness of our hybrid attention mechanism in the diffusion module. In both cases, our attention mechanism showed a performance improvement.

\begin{table}[!h] 
\centering
\setlength{\tabcolsep}{2mm}
\caption{Ablation studies of LoCI-DiffCom.}
\begin{tabular}{cccccc}
\hline
\multicolumn{2}{c}{Method} & \multirow{2}{*}{PSNR} & \multirow{2}{*}{SSIM} & \multicolumn{2}{c}{Dice} \\ \cline{1-2}  \cline{5-6} 
 LoCI     & Attention             &                       &                       & WM     & GM     \\ \hline
$\times 4$ & \checkmark     &\textbf{25.52}        &\textbf{0.845}         &0.656   &\textbf{0.650}\\

$\times 1$ & \checkmark     &24.60    &0.806       &0.605   &0.629\\
$\times 4$  &   &25.05        &0.830        &\textbf{0.661}   &0.647\\ 
$\times 1$  &   &24.20        &0.801       &0.600  &0.615\\

\hline
\end{tabular}
\label{Table:2}
\end{table}

\subsubsection{Analysis of Longitudinal Consistency Guidance}

The quality of the generated image could highly depend on guidance selection. We investigated the influence of different guidance selection strategies on the generated image by applying images at different ages as guidance to both conditional DDIM and our LoCI-DiffCom. As for generating image at 12 months of age, LoCI-DiffCom with any pair of guidance images imputed data with higher similarities to the ground truth in a highly robust manner (Fig. \ref{Fig:result}c). Conditional DDIM showed large instability, particularly with guidance far from the target age, where the generation quality was generally poor.

\begin{figure}[!t]
\centering
\includegraphics[width=0.7\textwidth]{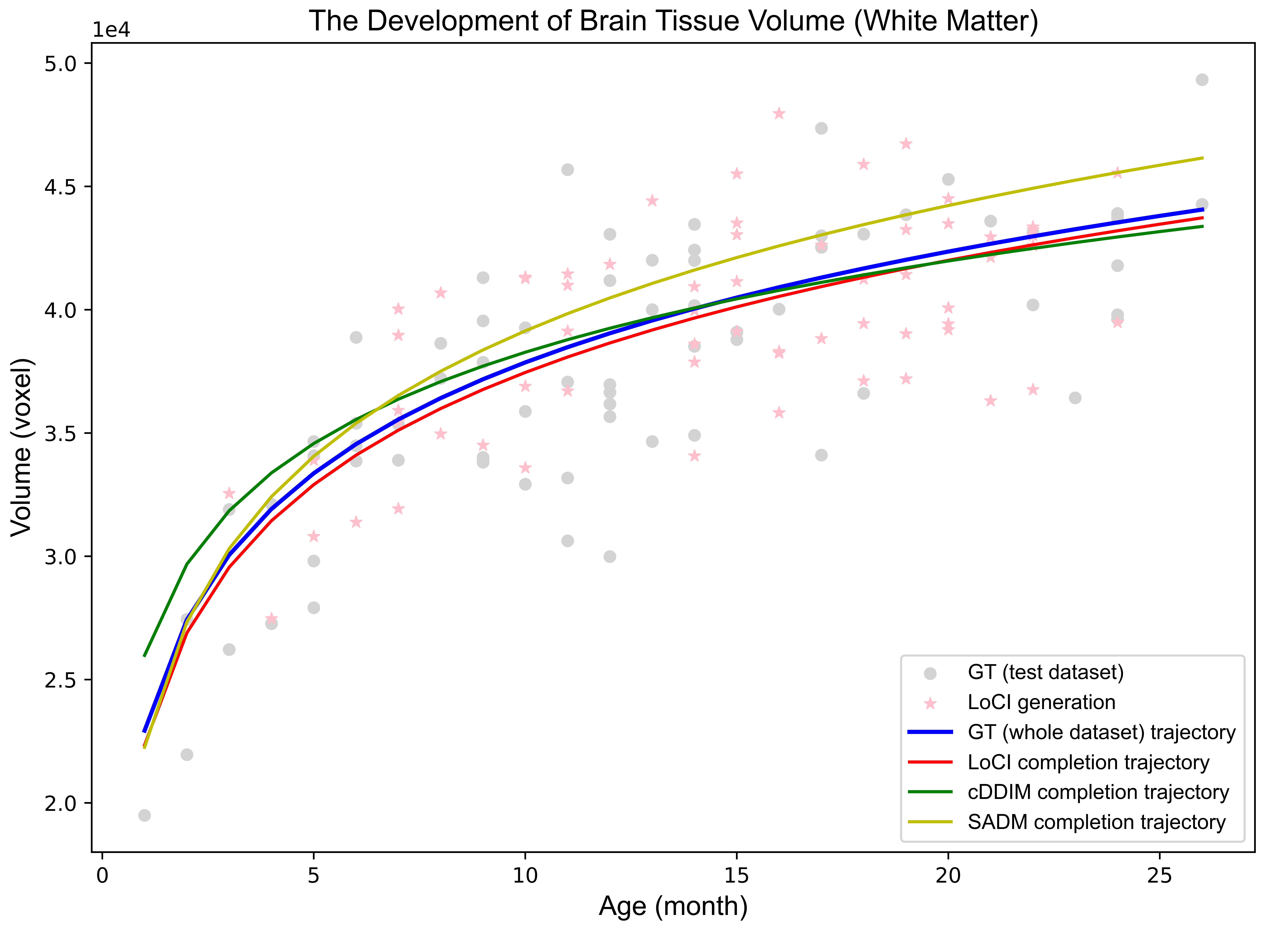}
\caption{The longitudinal growth trajectories of infant brain white matter volume.}
\label{Fig:trajectory_curve}
\end{figure}

\subsubsection{Performance on a Downstream Task: Delineating Developmental Trajectory after Data Completion.}
We further assessed the fidelity of data completion from a developmental neuroscience perspective by fitting the developmental trajectory of total white matter volume using a linear mixed-effect model with a log-linear function. Fig. \ref{Fig:trajectory_curve} demonstrates that the developmental trajectory completed by our method is closer to that of the ground truth, while SADM and conditional DDIM exhibit significant discrepancies with the ground truth trajectory.

\section{Conclusion}
We designed a novel conditional diffusion model, LoCI-DiffCom, aimed at completing missing infant brain images for better longitudinal studies. The proposed consistency-informed module effectively merges conditions from preceding and subsequent age time points and generates high-fidelity data that preserves longitudinal changes and individual variability. It is also highly stable in extreme scenarios where the guiding images are far from the target. Our approach offers a potential solution to complete missing data for more accurate developmental neuroscience studies.

%
%
%
%

\bibliographystyle{splncs04}
\bibliography{reference}

\end{document}